\documentclass{evn2004}
\usepackage{txfonts}
\usepackage{graphicx}
\begin{document}
\setcounter{page}{215}

   \title{Radio-loud and Radio-quiet X-ray Binaries: LSI+61$^{\circ}$303 in Context}
   \author{M. Massi}
   \institute{
Max Planck Institut f\"ur Radioastronomie, Auf dem H\"ugel 69,
D-53121 Bonn, Germany}

   \abstract{ The three basic ingredients - a spinning compact object,
an accretion disc and a collimated relativistic jet - make microquasars
a galactic scaled-down version of the radio-loud AGN.  That explains
the large interest attributed to this new class of objects, which up to
now consists of less than 20 members. Microquasars belong to the much
larger class of X-ray binary systems, where there exits a compact
object together with its X-ray emitting accretion disc, but the
relativistic jet is missing.  When does an X-ray binary system evolve
into a microquasar?  Ideal for studying such kind of a transition is
the periodic microquasar LS~I~+61$^{\circ}$303 formed by a compact
object accreting from the equatorial wind of a Be star and with more
than one event of super-critical accretion and ejection along the
eccentric orbit.  For ejections at periastron passage the relativistic
electrons suffer severe inverse Compton losses by upscattering the UV
photons of the Be star at high energy : At periastron passage Gamma-ray
emission has been observed, whereas radio outbursts have never been
observed in 20 years of radio flux monitoring.  For ejections displaced
from periastron passage the losses are less severe and radio outbursts
are observed.  The radio emission mapped on scales from a few AU to
hundreds of AU shows a double-sided relativistic ($\beta=0.6c$)
S-shaped jet, similar to the well-known precessing jet of
\object{SS~433}.  }

   \maketitle
\section{Introduction}
Since the beginning of the 1980s radio-galaxies, quasars, Seyferts, QSO
etc.  all are simply classified as AGNs (``Active Galactic Nuclei'')
because the ``energy-engine'' is thought to be the same: A
super-massive black hole accreting from its host galaxy.  AGNs having
radio-emitting lobes or jets are called radio-loud, the others are
called radio-quiet (Ulrich et al. 1997).

\begin{figure*}
   \centering
\includegraphics[width=15cm]{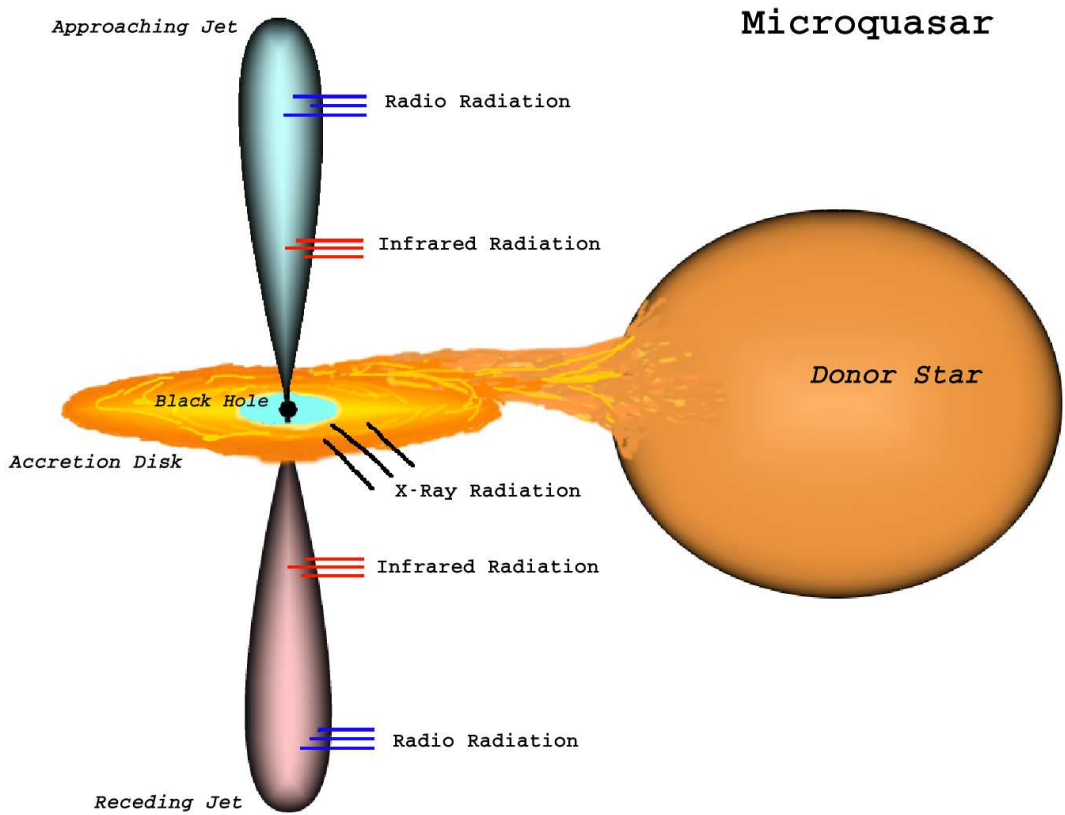}
   \caption{ The basic components of a microquasar - a spinning compact
object, an accretion disk and a collimated relativistic jet.  The
compact object, of a few solar masses, accretes from a normal star in
orbital motion around it.  The mass of the compact object can be
determined by studying the periodical shift of the optical spectral
lines of the normal star and establishes, if it is a neutron star or a
black hole.  The inner part of the disk emits X-rays. The inner radius,
three times the Schwarzschild radius, is a few tens of kilometers, the
outer radius a factor 10$^3$ larger (the figure is not to scale).  Due
to magneto-rotational instabilities part of the disk is propelled into
a relativistic jet, studied at high resolution with radio
interferometric techiques.  In some microquasars, like SS433 and LS
I+61$^{\circ}$303, the jet is precessing. If the precession causes the
jet to be aligned toward the earth the large variable doppler boosting
mimics the variability of Blazars and the microquasar in this case is
called microblazar.  }
\label{massi1}
    \end{figure*}

In an X-ray binary system the ``energy-engine'' is a compact object of
a few solar masses accreting from the companion star.  Up to now there
are almost 250 known X-ray binaries (Liu 2000).  Only a small
percentuage of them ($<$10\%) show evidence of a radio-jet and
therefore are radio loud applying the same definition as for the AGNs.
The radio loud X-ray binaries subclass (Fig.\ref{massi1}) includes
together with the microquasars --objects where high resolution radio
interferometric techniques have shown the presence of collimated jets
(Mirabel et al. 1992)-- also unresolved radio sources with a flat
spectrum.  This spectrum can arise from the combination of emission
from optically thick and thin regions of an expanding continuos jet
(Hjellming \& Johnston 1988; Fender 2004) as has been shown by the
discovery of a continuous jet for the flat-spectrum source Cygnus X-1
(Stirling et al. 2001).

Besides the first radio loud X-ray binary system SS433, discovered by
chance in 1979 (Margon 1984), the others have been discovered mostly in
the last ten years and now are considered as an ideal, nearby
laboratory for studying the processes of accretion and ejection around
black holes.

More generally speaking, the class of the X-ray binaries is formed by
stellar systems of two stars with very different natures: a normal star
(acting as mass donor) and a compact object (the accretor) that can be
either a neutron star or a black hole (White et al 1996).  The normal
star orbits around the compact object and therefore the infalling
material has some angular momentum ($J$), which prevents it from
falling directly into the accretor.  The stream of matter orbits the
compact object with a radius determined by $J$ and the mass of the
compact object ($M_x$).  The angular momentum is redistributed by the
viscosity: Some of the material takes angular momentum and spreads
outwards, whereas other material spirals inwards. In this way a disk is
created from the initial ring of matter (Longair 1994; King 1996).
Gradually, the matter drifts inwards until it reaches the last stable
orbit, The viscosity has two effects: Beside the transport of the
angular-momentum it also acts as a frictional force resulting in the
dissipation of heat.  The amount of friction depends on how fast the
gas orbits around the compact object.  It reaches its maximum at the
inner disk, where the matter is heated up to quite high temperatures
(tens of millions of degrees), producing the strong thermal X-ray
radiation giving the name ``X-ray binaries'' to these class of objects.

In the case of a low vertical magnetic field threading the disk the
plasma pressure dominates the magnetic field pressure and the
differentially rotating disk bends the magnetic field, which is
passively wound up (Meier et al. 2001).  Due to the compression of the
magnetic field lines the magnetic pressure may become larger than the
gas pressure at the surface of the accretion disk, where the density is
lower.  At this point the gas follows the twisted magnetic field lines,
creating two spinning flows.  These extract angular momentum from the
surface of the disk (magnetic braking) and enhance the radial
accretion.  The avalanching material further pulls the deformed
magnetic field with it and afterwards magnetic reconnection may happen
(Matsumoto et al. 1996).  The thickness of the disc is fundamental in
this magneto-rotational process, or better the extent of the poloidal
magnetic field frozen into the disc (Meier 2001; Meier et al. 2001;
Maccarone 2004).  No radio jet is associated at X-ray binaries in
High/soft states, where the X-ray spectrum is dominated by a
geometrically thin (optically thick) accretion disc (Shakura \& Sunyaev
1973).  Whereas, numerical results show a jet being launched from an
inner geometrically thick portion of the accretion disc existing
(coronal flow/ADAF) when the X-ray binaries are in their low/hard state
(Meyer et al. 2000; Meier 2001).

Among the X-ray binaries an ideal source to study the transition to the
microquasar phase is  LS~I~+61$^{\circ}$303 because it is the only 
known periodic microquasar.

\section {The LS~I~+61$^{\circ}$303 system}

LS~I~+61$^{\circ}$303 is the only object of this class showing
variations in X-rays (Leahy 2001), optical wavelenghts in both
continuum (Maraschi \& Treves 1981) and line radiation (Zamanov \&
Mart\'{\i} 2000; Liu et al. 2000) and at radio wavelenghts with a
period equal to the orbital one; the most accurate value of the orbital
period is that resulting from radio observations, equal to 26.496 days
(Gregory \& Taylor 1978; Taylor \& Gregory 1982; Gregory 2002).  The
fit performed on near infrared data by Mart\'{\i} and Paredes (1995)
produced high values for the eccentricity (e$\sim$ 0.7-0.8 ) confirmed
by optical observations (Casares et al. 2004).  The lower limit to $i$,
the angle formed by the axis of the orbit and the line of sight is
38$^{\circ}$ (Hutchings \& Crampton 1981; Massi et al. 2001; Massi
2004).  The phase at the periastron passage is $\Phi$=0.2 with the
phase referred to the time t$_0$=JD\,2443366.775, the date of the first
radio detection of the system (Gregory \& Taylor 1978).

Ultraviolet spectroscopy of LS~I~+61$^{\circ}$303 by Hutchings \&
Crampton (Hutchings \& Crampton 1981) indicates that the normal star is
a main sequence B0-B0.5 star (L$\sim 10^{38}$ erg sec$^{-1}$,
T$_{eff}\simeq 2.6~10^4$K).  The optical spectrum is that of a rapidly
spinning Be star.  Together with the usual high velocity (1000 km
s$^{-1}$) low density wind at high latitudes typical for OB stars, Be
stars have a dense and slow ($<$100 km~ s$^{-1}$) disk-like wind around
the equator (Waters et al. 1988).  Equatorial mass loss, due to an
interplay of the high rotation and of internal pulsations of the star,
is highly variable and in some cases periodical.
This is the case for LS~I~+61$^{\circ}$303, where periodical variations
of the mass loss from the Be star have been proved by a modulation of
H$\alpha$ emission line with a period of almost 4 years (Zamanov and
Mart\'{\i} 2000).

The most reliable method to determine the nature of the compact object
is the study, as usual in binary systems, of the changing radial
velocity of the normal companion during its orbit.  The amplitude
($K_c$) of the radial velocity variations and the period (P$_{orb}$) of
the system define a quantity, called the ``mass function'' (Charles \&
Wagner 1996), $f$, which depends on the inclination $i$ of the orbit,
the masses $M_X$ and $M$ of the accretor and its normal companion:
$$f ={P_{orb} K_c^3\over 2\pi G}={M_X^3 sin^3 i\over (M_X +M)^2} $$
where  $G$ is the gravitational constant.

Once the inclination, $i$, and the mass of the companion, $M$, are
 known one can solve for $M_X$.

 Rhoades \& Ruffini (1974) by taking the most extreme equation of state
that produces the maximum critical mass of a neutron star, established
the upper limit of 3.2 $M\odot$.  This absolute maximum mass provides a
decisive method of observationally distinguishing neutron stars from
black holes.  The problem with LS~I~+61$^{\circ}$303 is the large range
for the mass function allowed by the parameter uncertainties.  Optical
observations (Casares et~al.  2004) give the mass function $f$ in the
range $0.003<f<0.027$.  The upper limit for $f$, with i=38 and $M =
18\,M_\odot$ gives $M_{\rm X}<3.8\,M_\odot$.  Therefore the nature of
the accretor is still an open issue (Massi 2004).

\section {\bf Radio emission}

\begin{figure}[htb]
\centering
\resizebox{\hsize}{!}{\includegraphics[scale=0.15, angle=0]{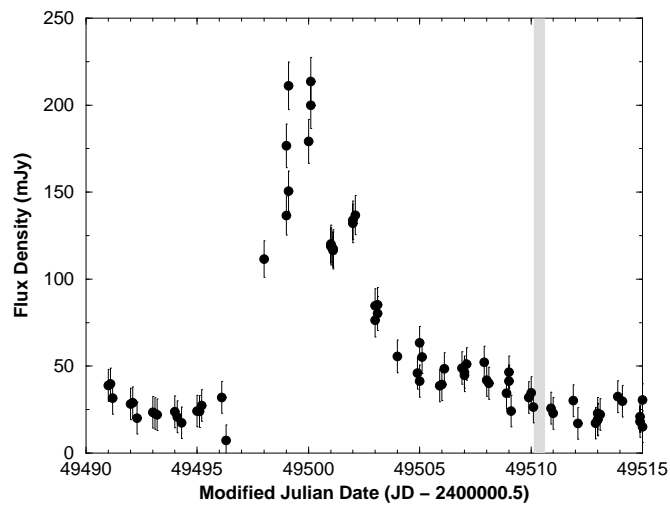}}
\caption{ Radio light curve of LS~I~+61$^{\circ}$303 obtained with the
GBI at 8.4~GHz (Strickman et~al. 1998). The shaded area indicates the
time interval during the EVN observation of Fig. 4 (figure in Massi et
al. 2001).}
\label{fig:flusso}
\end{figure}

The greatest peculiarity of LS I +61$^{\circ}$303 are its periodic
radio outbursts with P=26.496 days (Gregory 2002).  In
Fig.\ref{fig:flusso} a typical radio light curve is shown.  The decay
of the outburst agrees with that expected for an adiabatically
expanding cloud of synchrotron-emitting relativistic electrons (Taylor
\& Gregory 1984).  However, that model alone fails to fit the peaks at
different frequencies: the flat spectrum during the late portion of the
rise in flux density can be reproduced , if together with the adiabatic
expansion losses also a continuos ejection of particles lasting two
days is taken into account (Paredes et al 1991).

The amplitude of each outburst is not randomly varying, but itself
periodic with a periodicity of 4.6 years correlated with the mass loss
of the Be star (sect. 2) (Gregory 1999, 2002; Zamanov \& Mart\'{\i}
2000).  Also, the orbital phase $\Phi$ at which the outbursts occur is
modulated (Gregory et al. 1999) and varies in the interval 0.45--0.95
(Paredes et al. 1990) The $\Phi$ at periastron passage is
0.2. Therefore one of the fundamental questions concerning the periodic
radio outbursts of \object{LS~I~+61$^{\circ}$303} has been: Why are the
radio outbursts shifted with respect to the periastron passage?

\section {\bf The two peak accretion model}

\begin{figure}
   \centering
   \includegraphics[angle=-90,width=9.5cm]{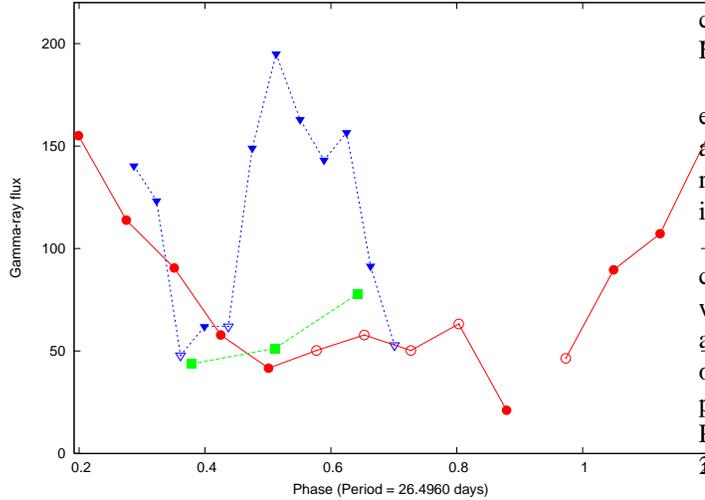}
      \caption{
EGRET data (Tavani et al. 1998) folded with the orbital/radio period
$P=26.4960$ days.  The plot begins at periastron passage
$\Phi\simeq$0.2 and shows the follow-up of the gamma-ray emission along
one full orbit. At epoch 2450334JD (i.e. circles in the plot, with
empty circles indicating upper limits) the orbit has been well sampled
at all phases: A clear peak is centered at periastron passage 0.2 and
1.2.  At a previous epoch (2449045JD; triangles in the plot, with empty
triangles indicating upper limits) the sampling is uncomplete, still
the data show an increase toward ($\Phi \simeq $0.3) periastron passage
and a peak at $\Phi\simeq$0.5. The 3 squares refer to a third epoch
(2449471JD).  The emission is suggested to be produced via inverse
Compton scattering of stellar photons (Taylor et~al. 1992, 1996) and of
synchrotron photons (Bosch-Ramon \& Paredes 2004) by the relativistic
electrons of the jet. The ejection is predicted to occur twice along
the orbit, one always at periastron passage and the second at a varing
orbital phase (Mart\'{\i} and Paredes 1995).  Relativistic electrons
ejected at $\Phi$=0.2 suffer severe Compton losses: radio outbursts
never occur at periastron.  Radio outbursts occur in the orbital phase
interval 0.45-0.95 (Paredes et al. 1990).  The gamma-ray peak at
$\Phi\simeq0.5$ could be associated to a such second ejection.
However, no radio data are available to calculate energy budget/losses
of the relativistic electrons (figure in Massi et al. 2004b).  }
         \label{massi3}
   \end{figure}

In order to explain the association between LS~I~+61$^{\circ}$303 and
the gamma-ray source 2CG 135+01/3EG J0241+6103, Bosch-Ramon and Paredes
(2004) have proposed a numerical model based on inverse Compton
scattering.  The relativistic electrons in the jet are exposed to
stellar photons (external Compton) as well as to synchrotron photons
(synchrotron self Compton).  The model considers accretion variations
along the orbit and predicts a gamma-ray peak at periastron passage
where the accretion is higher.  EGRET data show indeed a gamma-ray peak
at $\Phi$=0.2 (Fig.\ref{massi3}) and catastrophic inverse Compton
losses might explain the absence of radio emission at periastron.
Therefore, to explain the observed periodic radio outbursts in the
phase interval 0.45--0.95 a second accretion/ejection event must occur.

Taylor et al. (1992) and Mart\'{\i} \& Paredes (1995) have shown that
for accretion along an eccentric orbit the accretion rate $\dot{M}
\propto {\rho_{\rm wind}\over v_{\rm rel}^3}$, (where $\rho_{\rm wind}$
is the density of the Be star wind and $v_{\rm rel}$ is the relative
speed between the accretor and wind) develops two peaks: the highest
peak corresponds to the periastron passage (highest density), while the
second peak occurs when the drop in the relative velocity $v_{\rm rel}$
compensates the decrease in density (because of the inverse cube
dependence).  Mart\'{\i} \& Paredes (1995) have shown that during both
peaks the accretion rate is above the Eddington limit and therefore one
expects that matter is ejected twice within the 26.496 days interval.
Mart\'{\i} \& Paredes have found that variations of Be star wind
velocity produce a variation in the orbital phase of the second peak.
At this second accretion peak the compact object is far enough away
from the Be star, so that the inverse Compton losses are small and
electrons can propagate out of the orbital plane.  Then an expanding
double radio source should be observed, which in fact has been observed
by VLBI and MERLIN.

\section{A precessing  jet}

The first VLBI observation resolving the source, made by
Massi and collaborators (1993; $\Phi$=0.74)
\begin{figure}
\centering
\includegraphics[width=8.5cm]{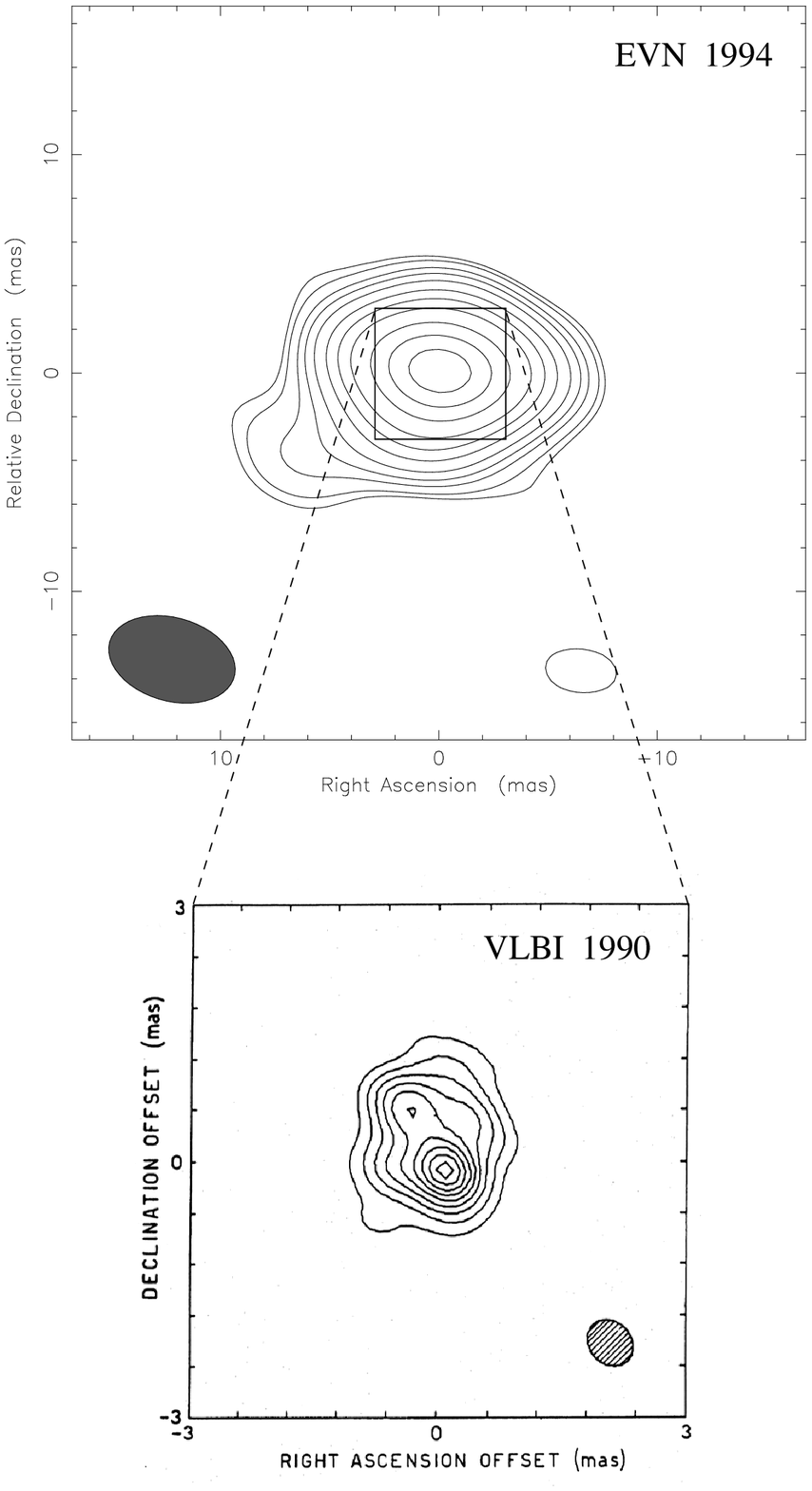}
\caption{ Bottom: VLBI observation of LS~I~+61$^{\circ}$303 at 6cm.
 The telescopes involved were those at Effelsberg (Germany), Westerbork
 (Neth erlands), Medicina (Italy), Onsala (Sweden) and VLA (New Mexico,
 USA).  The observation, lasting 14 hours, was performed during a slow
 decay of a large ($>250$mJy) outburst. The beam is 0.6 x 0.5 mas. The
 peak flux density is 36 mJy/beam. The lowest contour is 15$\%$ of the
 peak and the increment is of 10$\%$. Top: EVN uniform weighted map of
 LS~I~+61$^{\circ}$303 at 6 cm.  The telescopes involved were those at
 Effelsberg , Medicina , Noto (Italy) and Onsala.  The contours are at
 $-$3, 3, 4, 6, 8, 11, 16, 22, 30, 45, 60 and 75 times the r.m.s. noise
 of 0.28~mJy~beam$^{-1}$. The filled ellipse in the bottom-left corner
 represents the FWHM of the synthesized beam, which is
 5.9~mas$\times$3.8~mas at a P.A.  of 74.2$^{\circ}$ (figure in Massi
 et al. 2001).}
\label{fig:vlbi+evn}
\end{figure}
revealed a complex morphology (Fig.\ref{fig:vlbi+evn}~bottom) not easy
 to interprete.  Two components (at P.A.$\simeq$30$^{\circ}$) are
 separated by 0.9 mas, corresponding to 1.8 AU at a distance of 2 kpc.
 The two components are inside an extended and sensibly rotated
 structure at P.A.$\simeq$ 135$^{\circ}$).  What is the nature of this
 complex morphology ?  Is the envelope an older expanding jet,
 previously ejected, because of precession, at another angle?  Taylor
 and collaborators (2000; $\Phi$=0.69) performing VLBI observations in
 combination with the HALCA orbiting antenna mapped a curved structure
 of 4 mas (8 AU) reminiscent ''of the precessing radio jet seen in
 SS433''.

EVN observations (Massi et al 2001; $\Phi$=0.92) at a scale of up to
tens of AU show an elongation clearly in one direction without any
ambiguity (see Fig.\ref{fig:vlbi+evn}-Top).  The observed flux density
of the approaching ($S_{\rm a}$) and the receding ($S_{\rm r}$) jet are
a function of $\theta$, the angle between the jet and the line of
sight, by the Doppler factor: $\delta_{\rm a,r}=[ \Gamma (1 \mp
\beta\cos\theta)]^{-1}$, where $\Gamma=(1-\beta^2)^{-1/2}$ is the
Lorentz factor and $\beta~c$ the jet velocity (Mirabel \& Rodriguez
1999).  The observed flux density of the approaching jet will be
boosted and that of the receding jets de-boosted as $S_{\rm a,r}=S
\delta_{\rm a,r}^{k-\alpha}$, where $\alpha$ is the spectral index of
the emission ($S_{\nu}\propto \nu^{+\alpha}$) and $k$ is 2 for a
continuous jet and 3 for discrete condensations.  That creates an
asymmetry between the two jet components. The receding attenuated jet
can even dissapear because of the sensitivity limit of the image. In
this case the jet appears only on one side as it is the case for the
EVN image.

\begin{figure*}
   \centering
   \includegraphics[width=12cm]{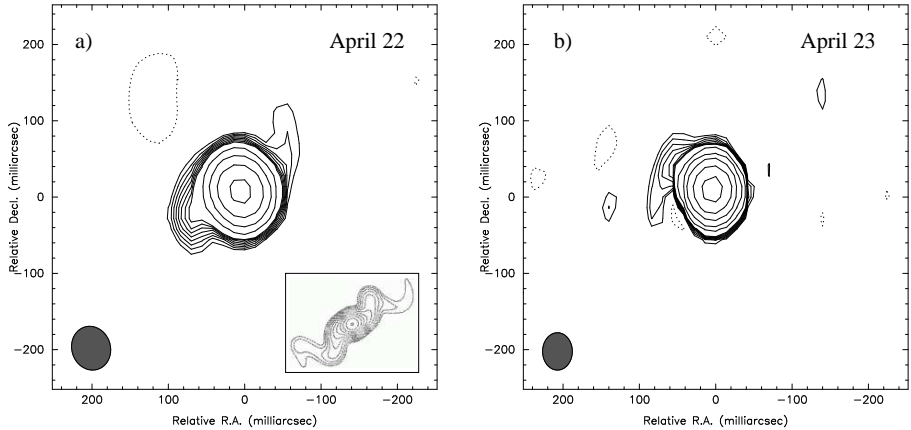}
      \caption{ {\bf a)} MERLIN image of \object{LS~I~+61$^{\circ}$303}
at 5~GHz obtained on 2001 April 22. North is up and East is to the
left. The synthesized beam has a size of $51\times58$~mas, with a PA of
17\degr. The contour levels are at $-$3, 3, 4, 5, 6, 7, 8, 9, 10, 20,
40, 80, and 160$\sigma$, being $\sigma$=0.14~mJy~beam$^{-1}$. The
S-shaped morphology strongly recalls the precessing jet of
\object{SS~433}, whose simulated radio emission (Fig.~\ref{massi6}b in
Hjellming \& Johnston \cite{hjellming88} ) is given in the small
box. {\bf b)} Same as before but for the April 23 run. The synthesized
beam has a size of $39\times49$~mas, with a PA of $-$10\degr. The
contour levels are the same as those used in the April 22 image but up
to 320$\sigma$, with $\sigma$=0.12~mJy~beam$^{-1}$ (figure in Massi et
al. 2004).  }
         \label{massi5}
   \end{figure*}

The first of two consecutive MERLIN observations (Massi et al. 2004;
$\Phi$=0.68) shows a double S-shaped jet extending to about 200~AU on
both sides of a central source (Fig.\ref{massi5}a).  The receding jet is
attenuated but still above the noise limit of the image.

The precession suggested from the first MERLIN image becomes evident in
the second one ($\Phi$=0.71), shown in Fig.\ref{massi5}b, where a new
feature is present oriented to the North-East at a position angle (PA)
of 67$\degr$.  It is likely that the morphology of the source is
S-shaped because the only visible jet appears indeed bent.  The
Northwest-Southeast jet of Fig.\ref{massi5}a has a PA=124$\degr$.
Therefore a quite large rotation has occurred in only 24 hours.

The appearance of successive ejections of a precessing jet with
ballistic motion of each ejection is a curved path, that depending on
the modality of the expansion and therefore on the adiabatic losses
seems to be a ``twin-corkscrew'' or a simply S-shaped pattern
(Hjellming \& Johnston 1988; Crocker et al. 2002) Can we distinguish in
our data the single ejections in ballistic motion?
 
\begin{figure*}
   \centering
  \includegraphics[width=\textwidth]{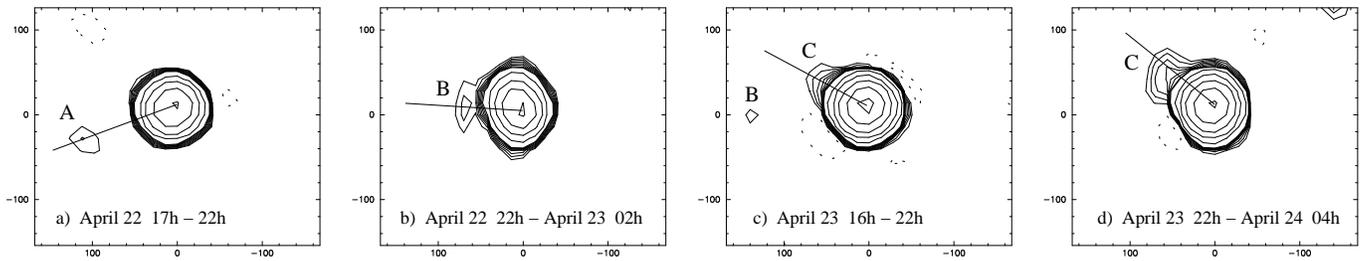}
\caption{ MERLIN images of LS~I~+61$^{\circ}$303 of 2001 April 22 and
 April 23. The data set of each epoch has been split into two blocks
 (i.e. Fig.~\ref{massi5}a is the combination of the first two blocks:
 6a and 6b; Fig. \ref{massi5}b a combination of 6c and 6d).  A
 convolving beam of 40~mas has been used in all images for better
 display. The first contour represents the $3\sigma$ level in all
 images except for c), where we start from the $2\sigma$ level to
 display the faint B component.  The rms noises are
 $\sigma$=0.13~mJy~beam$^{-1}$, $\sigma$=0.20~mJy~beam$^{-1}$,
 $\sigma$=0.13~mJy~beam$^{-1}$, and $\sigma$=0.15~mJy~beam$^{-1}$,
 respectively.  The PA of the ejections is indicated by a bar (figure
 in Massi et al. 2004).}
         \label{massi6}
   \end{figure*}

We have split the MERLIN data of each epoch into blocks of a few hours
and created separate images (Fig.~\ref{massi6})
(i.e. Fig.~\ref{massi5}a is the combination of the first two blocks:
\ref{massi6}a and \ref{massi6}b; Fig.~\ref{massi5}b a combination of
\ref{massi6}c and \ref{massi6}d).  We see that the Eastern bent
structure present in Fig.~\ref{massi5}a is the result of a combination
of an old ejection A (Fig.~\ref{massi6}a), already displaced 120~mas
from the core, and a new ejection B (Fig.~\ref{massi6}b). After 19
hours (Fig.~\ref{massi6}c) the feature B is reduced to $2\sigma$ and a
new ejection C, at a different PA with respect to B, appears. In
Fig.~\ref{massi6}d, 6 hours later, little rotation of the PA is
compatible with $\Delta$ PA$_{(\rm B-\rm C)}/ 3$ of the previous image.

From the maps it is evident that the projection of the jet on the sky
plane is changing, but how much is the variation of $\theta$?  If the
jet velocity is the same for all ejections, the change of the ratio
${S_{\rm a}\over{S_{\rm
r}}}=\left({1+\beta\cos\theta\over1-\beta\cos\theta}
\right)^{k-\alpha}$ is due to variations of $\theta$.  Adopting values
of $k=2$, $\alpha$=-0.5 and the a value of $\beta=0.6$ we obtain:
$\theta_{\rm A} < 90\degr$, $\theta_{\rm B}< 80\degr$ and for the C
ejection in Fig.~\ref{massi6}c, $\theta_{\rm C} < 68 \degr$.

\section{Conclusions}

The main points of our review on the characteristics of the source
LS~I~+61$^{\circ}$303 in the frame of the two-peak accretion/ejection
model are:

\begin{enumerate}
\item
It is still an open issue whether the compact object in this system is
a neutron star or a black hole.  In fact, taking into account the
uncertainty in inclination, mass of the companion and the mass
function, the existence of a black hole cannot be ruled out.

\item
The radio jet at a scale of hundreds of AU quite strongly changes its
morphology in short intervals (within 24 hours), evolving from an
initial double-sided jet into an one-sided jet. This variation
corresponds to a reduction of more than 10$^{\circ}$ in the angle
between the jet and the line of sight. This new alignment severely
Doppler de-boosts the counter-jet.  Further observational evidence for
a precessing jet is recognizable even at AU scales.

\item
The same population of relativistic electrons emitting
radio-synchrotron radiation may upscatter - by inverse Compton
processes - ultraviolet stellar photons and produce gamma-ray emission.
For ejections at the periastron passage gamma-ray flares are expected,
but because of severe Compton losses no radio flares, as indeed the
data seem to indicate.

\end{enumerate}

We conclude that as precession and variable doppler boosting are the
causes of the rapid change in the radio-morphology, precession and
variable doppler boosting are likely to produce gamma-ray variations at
short time scales.  The amplification due to the Doppler factor for
Compton scattering of stellar photons by the relativistic electrons of
the jet is $\delta^{3-2\alpha}$ (where $\alpha<0$), and therefore
higher than that for synchrotron emission, i.e. $\delta^{2-\alpha}$
(Kaufman Bernad\'o et~al. 2002).  LS~I~+61$^{\circ}$303 becomes
therefore the ideal laboratory to test the recently proposed model for
microblazars with INTEGRAL and MERLIN observations now and by AGILE and
GLAST in the future.

\begin{acknowledgements}

It is a pleasure to thank Karl Menten and J\"urgen Neidh\"ofer for
careful reading of the manuscript and valuable comments and
discussions.

\end{acknowledgements}

\end{document}